\pdfoutput=1
\documentclass[pdftex,12pt]{JHEP3}
\usepackage{amsfonts,amsbsy,latexsym,shadow,fancybox,amssymb,amscd,amstext,amsmath}
\usepackage[pdftex]{epsfig}
\pdfoutput=1

\renewcommand{\AmS}{{\protect\the\textfont2A\kern-.1667em\lower.5ex\hbox{M}\kern-.125emS}}
 \def\be{\begin{equation}}
\def\ee{\end{equation}}
\def\bea{\begin{eqnarray}}
\def\eea{\end{eqnarray}}
\def\pd{\partial}
\def\a{\alpha}
\def\b{\beta}

\def\d{\delta}
\def\m{\mu}
\def\n{\nu}
\def\t{\tau}

\def\l{\lambda}

\def\r{\rho}
\def\s{\sigma}
\def\e{\epsilon}

\def\bg{\bar{g}}
\def\bn{\bar{\nabla}}

\def\cR{\cal R}

\def\bg{\bar{g}}
\def\bs{\bar{\s}}

\def\bn{\bar{\nabla}}
\def\bi{\begin{itemize}}
\def\ei{\end{itemize}}
\def\bn{\bar{\nabla}}

\date{ 2011} \preprint{IFT-UAM/CSIC-12-116;FTUAM-12-117}
\title{No Conformal Anomaly in Unimodular Gravity.} \author{Enrique \'Alvarez $\star$ $\bullet$ and Mario Herrero-Valea $\bullet$ \\ $\star$ Physics Department. Theory Unit~~CERN~1211~Gen\`eve 23, Switzerland \\$\bullet$  Instituto de F\'{\i}sica Te\'orica
UAM/CSIC and Departamento de F\'{\i}sica Te\'orica \\ Universidad
Aut\'onoma de Madrid, E-28049--Madrid, Spain \\ E-mail: \email{enrique.alvarez@uam.es $\quad$ mario.herrero@estudiante.uam.es }}
\abstract{The conformal invariance of unimodular gravity survives quantum corrections, even in the presence of conformal matter. Unimodular gravity can actually be understood as a certain truncation of the full Einstein-Hilbert theory, where in the Einstein frame the metric tensor enjoys unit  determinant. Our result is compatible with the idea that the corresponding restriction  in the functional integral is  consistent as well.
}
\begin{document}

{\vskip 1cm}
\newpage
\vskip 1cm

\vskip 1cm


\newpage
\tableofcontents


\setcounter{page}{1}
\setcounter{footnote}{1}
\tableofcontents

\newpage
\section{Introduction.}

A radical approach towards explaining why (the zero mode of) the vacuum energy seems to violate the equivalence principle (the {\em active cosmological constant problem}) is just to eliminate the direct coupling in the action between the potential energy and the gravitational field \cite{AlvarezF}. This leads to consider unimodular theories, where  the metric tensor is constrained to be unimodular 
\be
g_E\equiv \left|\text{det} g^E_{\m\n}\right|=1
\ee
 in the Einstein frame. This equality only stands in those reference frames obtained from the Einstein one by an area preserving diffeomorphism. Those are by definition the ones that enjoy unit jacobian, and $g$ is a singlet under them.
 \par
 We shall represent the absolute value of the determinant of the metric tensor in an arbitrary frame as $g$ instead of $|g|$ in order to simplify the corresponding formulas. We work in arbitrary dimension $n$ in order to be able to employ dimensional regularization as needed.
The simplest nontrivial such unimodular action \cite{AlvarezF}  reads
\bea
&&S_U\equiv - M^{n-2}\int d^n x ~R_E~+~ S_{matt}=\nonumber\\
&&- M^{n-2}\int d^n x~g^{1\over n}\left(~R+{(n-1)(n-2)\over 4 n^2} {g^{\m\n}\nabla_\m g~\nabla_\n g\over g^2}\right)+S_{matt}
\eea
where the n-dimensional Planck mass is related to the n-dimensional Newton constant through
\be
M^{n-2}\equiv {1\over 16 \pi G}.
\ee
and $S_{matt}$ is the matter contribution to the action.
\par
This theory is conformally (Weyl) invariant under
\be
\tilde{g}_{\m\n}=\Omega^2(x)g_{\m\n}(x)
\ee
(the Einstein metric is inert under those) as well as under area preserving (transverse) diffeomorphisms, id est, those that enjoy unit jacobian, thereby preserving the Lebesgue measure. We shall speak always of {\em conformal invariance} in the above sense.
\par
   The aim of this paper is to explore whether this gauge symmetry is anomalous or survives when one loop quantum corrections are taken into account. The result we have found is that, given the fact that  this theory can be thought as a partial gauge fixed sector of a conformal upgrading of General Relativity, there is no conformal anomaly for unimodular gravity, even when conformal matter is included.  
\par
Other interesting viewpoints on the cosmological constant from the point of view of unimodular gravity are \cite{Ng} \cite{Smolin}. In this last reference Smolin suggested the absence of conformal anomaly for related theories.

We will proceed as follows. First, we will define a more general scalar-tensor theory by introducing a spurion field $\sigma$. This theory is diffeomorphism as well as conformal invariant and unimodular gravity is no more than a partial gauge fixed sector of it. This happens to be, also, the same theory that t'Hooft proposed \cite{Hooft} in order to solve some special issues of black hole complementarity. Consequently, in section \ref{t'Hooft} we will explore t'Hooft's approach in order to obtain the divergent part of the one-loop effective action of such theory. In section \ref{invariancia}, however, we will show how the scalar-tensor action can be written in a more useful and manifestly conformal invariant form and we will use it to easily compute the one-loop gravitational counterterm in section \ref{one-loop}. Our result is not only consistent with t'Hooft¡s computations but also shows in a very clear way how the anomaly vanishes.
\newpage
\section{A more general scalar-tensor theory}
It is technically quite complicated to gauge fixing a theory invariant under area preserving (transverse) diffeomorphisms only, because the theory is {\em reducible}, id est,  the corresponding gauge parameters are not independent. This usually demands a huge ghost sector \cite{HenneauxT}. In order to avoid these, presumably physically irrelevant intricacies , it proves convenient to introduce a new theory which enjoys diffeomorphism invariance and that is such that the unimodular theory is a partial gauge fixing of it. This is easily achieved by introducing a {\em compensating field}, $C(x)$, defined so that
\be
\label{fixing}g(x) C^2\equiv e^{{2n\over \sqrt{(n-1)(n-2)}}\s(x)}
\ee
transforms as a true scalar. This diffeomorphism invariant theory is still Weyl invariant provided that
\be
e^{{2n\over \sqrt{(n-1)(n-2)}}\tilde{\s}(x)}=\Omega^{2n}~e^{{2n\over \sqrt{(n-1)(n-2)}}\s(x)}.
\ee
Id est, the composite exponential field has got {\em conformal weight} $-2n$. In general, a conformal tensor of conformal weight $-\l$ behaves under conformal transformations as
\be
\d T= \l ~T
\ee
\par
At the linear level, $\Omega(x)\equiv 1+\omega(x)$, the spurion field $\s$ transforms with the gauge parameter like a true  {\em Goldstone boson} does. 
\be
\d\s=\sqrt{(n-1)(n-2)}~\omega
\ee
This result conveys the fact that the spurion is nothing else than the {\em dilaton}.
 The unimodular theory of our interest is recovered when the partial {\em unitary} gauge 
\be
C=1
\ee
is chosen. The residual gauge symmetries are then the area preserving diffeomorphisms as well as Weyl invariance.

Let us be quite explicit on this point. Under a diffeomorphism
\be
\d x^\m=\xi^\m
\ee
the compensating field behaves as
\be
\d C=\pd_\l \xi^\l C-\xi^\l\pd_\l C
\ee
whereas it is a singlet under conformal transformations. Under finite transformations
\be
C^\prime(x^\prime)=C(x).\text{det}{\pd x^\prime\over \pd x}.
\ee
To reach the gauge $C=1$ starting from a non-vanishing $C\neq 0$ it is then enough to choose
\be
\text{det}\left({\pd x^\prime\over \pd x}\right)={1\over C(x)}
\ee
The gauge $C=1$ means that
\be
e^{{2n\over\sqrt{(n-1)(n-2)}}\s(x)}=g(x).
\ee

\par
The new action is then written as
\bea
&&S\equiv \int d^n x~ \sqrt{g}\bigg\{~e^{-\sqrt{n-2\over n-1}\s}\left[ - M^{n-2}\left(~R+~ g^{\m\n}\nabla_\m \s~\nabla_\n \s\right)+{1\over 2}~g^{\m\n}\nabla_\m\Phi\nabla_\n\Phi\right]\nonumber\\
&&-e^{-{n\over\sqrt{(n-1)(n-2)}}\s}~V(\Phi)\bigg\}.
\eea
This action is conformal invariant as well as diffeomorphism invariant with all fields transforming as indicated above.
\par

The spurion $\s(x)$ corresponds to a conformal rescaling of the metric, and behaves as a ghost (its kinetic energy term has got the wrong sign). It is standard, since the work \cite{Gibbons} to perform its functional integral over imaginary values of the field. We shall keep an open mind on this issue for the time being.
\par

The canonically normalized field is
\be
\phi_g\equiv - 2\sqrt{2} M^{n-2\over 2}~\sqrt{n-1\over n-2}~e^{-{1\over 2}~\sqrt{n-2\over n-1}\s}.
\ee
The old gauge $C=1$ now reads

\be
\phi_g+2^{3\over 2} M^{n-2\over 2}\sqrt{n-1\over n-2} g^{-{n-2\over 4 n}}=0.
\ee

In terms of $\phi_g$ the action is
\bea
&&S_{ST}=\int d^n x \sqrt{g}~\bigg\{~-{n-2\over 8(n-1)}~R~ \phi_g^2-{1\over 2}g^{\m\n}\nabla_\m\phi_g\nabla_\n \phi_g+\nonumber\\
&&{n-2\over 8(n-1) M^{n-2}}\phi_g^2 {1\over 2} (\nabla\Phi)^2-\left(-1\right)^{2n\over n-2}\left({n-2\over 8(n-1)}\right)^{n\over n-2}{1\over M^n}\phi_g^{2 n\over n-2}V(\Phi)\bigg\}.
\eea
It is instructive to study in detail how the equations of motion (EM) of the scalar-tensor theory reduce to the unimodular ones in the unitary gauge.
Indeed,

\be
\left.{\d S_U\over \d g_{\m\n}}={\d S_{ST}\over \d g_{\m\n}}+{\d S_{ST}\over \d \phi_g}{\d \phi_g\over \d g_{\m\n}}\right|_{\phi_g=-2^{3\over 2} M^{n-2\over 2}\sqrt{n-1\over n-2} g^{-{n-2\over 4 n}}}.
\ee

This conveys the fact that the scalar-tensor EM imply the unimodular EM, whereas the converse assertion is untrue: the unimodular EM do not imply the scalar-tensor ones. The unimodular theory is a subsector of the more general scalar tensor theory, as stated in the title of the present section.
\par
In this scalar-tensor theory it is possible to go to the  Einstein frame through
\be
g_{\m\n}=2^{6\over n-2}M^2\left({n-1\over n-2}\right)^{2\over n-2}\phi_g^{-{4 \over n-2}}g^E_{\m\n}.
\ee
This metric $g^E_{\m\n}$ is a conformal singlet; that is, it remains invariant under Weyl transformations. 
In the gauge $C=1$ we go round the whole circle and the metric in  Einstein's frame is unimodular,
\be
\{C=1\}\Rightarrow\quad g_E=1.
\ee
\par
It is also possible here to define, again following \cite{Gibbons}\cite{Hooft},
\be
i \text{log}~\phi_g\equiv \eta.
\ee

\par
In the following we will just work with the gravitational sector, since as we will show later, inclussion of matter will not change any of our conclusions. Thus, in the rest of this text we will forget about the scalar field.

\section{ 't Hooft's approach: effective action after integrating the conformal factor.}\label{t'Hooft}
The gravitational piece of the previous Lagrangian is identical to the one proposed by 't Hooft in \cite{Hooft} in order to solve conceptual problems of black holes. For this purpose it is essential to integrate first over the scalar field in such a way as to get a conformally invariant theory of gravity. The divergent part of the functional integral over the scalar field can be easily computed after it is  conveniently rotated to imaginary values, as advertised earlier
\be
 e^{-{1\over 2}\sqrt{n-2\over n-1}\s(x)}\equiv 1+i \a(x)\quad \a\in\mathbb{R}
\ee
and the result of this integration over ${\cal D}\a$ can be expressed in terms of the Weyl tensor. Let us remind its origin.
The {\em Schouten tensor} is defined as
\[
A_{\a\b}\equiv {1\over n-2}\left(R_{\a\b}-{1\over 2(n-1)} R g_{\a\b}\right)
\]
and the Weyl tensor reads
\[
W_{\a\b\m\n}\equiv R_{\a\b\m\n}+\left(A_{\b\m}g_{\a\n}+A_{\a\n}g_{\b\m}-A_{\b\n}g_{\a\m}-A_{\a\m}g_{\b\n}\right).
\]
Under conformal transformations, it transforms as a conformal tensor of weight $\l=-1$:
\be
\tilde{W}_{\a\b\m\n}\equiv e^{2\s}~W_{\a\b\m\n}.
\ee
So its square has got weight $\l=2$ in such a way that
\be
|g|^{2/n} W^{\m\n\r\s}W_{\m\n\r\s}=|g|^{2/n}\left(R_{\m\n\r\s} R^{\m\n\r\s}-2 R_{\m\n} R^{\m\n}+{1\over 3} R^2\right)
\ee
is pointwise invariant (but behaves as a true scalar in four dimensions only).
\par
The Weyl tensor  vanishes identically in low dimension $n=2$ and $n=3$ and a space with $n\geq 4$ is conformally flat iff $W=0$. In that case, the claim is that the counterterm reads
\be
L_{div}=-{\sqrt{g}\over 480\pi^2 \left(n-4\right)} W_{\m\n\r\s}W^{\m\n\r\s}.
\ee
This functional behavior (barring the coefficient) could have been predicted from the fact that the result theory had to be pointwise conformal invariant. It can also be written as
\be
L_{div}={\sqrt{g}\over 960\pi^2 \left(n-4\right)} \left(R_{\m\n}R^{\m\n}-{1\over 3} R^2\right).
\ee
The second expression is easily obtained assuming that the Euler topological invariant vanishes, id est
\be
\int d^4 x \sqrt{g}\left(R_{\m\n\r\s}R^{\m\n\r\s}-4 R_{\m\n} R^{\m\n} - R^2\right)=0.
\ee
It is perhaps worth remarking that the  quantity
\be
\d \left(g^{2\over n} W_{\m\n\r\s}\ W^{\m\n\r\s}\right)=0
\ee
which is invariant under area preserving diffeomorphisms only, enjoys pointwise conformal invariance {\em in any dimension}, when the power of the determinant is determined in order to enhance area preserving diffeomorphisms  to the full group of diffeomorphisms, the resulting expression is conformal invariant only in dimension $n=4$
\be
\d \left( \sqrt{g} W_{\m\n\r\s}\ W^{\m\n\r\s}\right)=-{4-n\over 2n}~2 n \omega(x)~ \sqrt{g}~W_{\m\n\r\s}\ W^{\m\n\r\s}~.
\ee
This fact, first noticed by Duff \cite{Duff} leads to the understanding of the standard conformal anomaly in dimensional regularization through finite remainders coming from the $\e~{1\over \e}$ cancellation.

\section{Conformal invariance}\label{invariancia}

Instead of working with the scalar-tensor theory in the form we just obtained, let us clarify its physical content by defining the following vector field

\begin{align}
W_{\mu}\equiv \frac{1}{n-2}e^{-\sqrt{\frac{n-2}{n-1}}\sigma}\nabla_{\mu}e^{\sqrt{\frac{n-2}{n-1}}\sigma}=\frac{1}{\sqrt{(n-2)(n-1))}}\nabla_{\mu}\sigma
\end{align}

which under conformal transformations behaves as an abelian gauge field
\begin{align}
W^{'}_{\mu}=\Omega^{-1}\nabla_{\mu}\Omega + W_{\mu}.
\end{align}
This fact encodes a deep meaning, namely that in general we should be always able to construct a pointwise invariant conformal theory from a non-invariant one by adding interactions with this gauge field in a similar way as it is done in a Yang-Mills theory to implement local invariance under $SU(N)$ to the fermionic matter. This is precisely the situation we have in the unimodular theory (which is more clear when described through this more general scalar-tensor theory), which naively, and forgetting for the moment the implications of the $C=1$ partial gauge fixing, is no more than an upgrading of Einstein-Hilbert theory into a conformal invariant one, so it has to be possible to rewrite it just as General Relativity coupled to this $W_{\mu}$ field. 

Thus, let us start as usual by defining a gauge covariant derivative by meanings of the gauge connection, which upgrades the riemmanian connection to
\begin{align}
\Gamma(W)^{\mu}_{\nu\rho}=\Gamma^{\mu}_{\nu\rho}-\delta^{\mu}_{\nu}W_{\rho}-\delta^{\mu}_{\rho} W_\n +g_{\nu\rho}W^{\mu}
\end{align}
which allows us to define a conformal (as well as diffeomorphism) covariant derivative by
\begin{align}
D_{\mu}T=\nabla_{\mu}^{\Gamma(W)}T+\lambda W_{\mu}T
\end{align}
where $-\lambda$ is the conformal weight of the tensor T and $\nabla_{\mu}^{\Gamma(W)}$ states for the derivative defined through the Weyl connection $\Gamma (W)$.

The important fact that arises here is that even if this Weyl connection is not a metric one, all dynamical quantities can however be canonically constructed just by defining a new metric in such a way that
\begin{align}
G_{\alpha\beta}=e^{-\frac{2\sigma}{\sqrt{(n-2)(n-1)}}}g_{\alpha\beta}\; \longrightarrow \; \Gamma(W)^{\mu}_{\nu\rho}\left[ g_{\alpha\beta}\right]=\Gamma^{\mu}_{\nu\rho}\left[ G_{\alpha\beta}\right]
\end{align}
which enjoys all expected properties.

So at this point things are straightforward and we can compute naive Weyl invariant (once proper integration measure is provided) geometrical quantities out of the $D_{\m}$ derivative, such as the Riemman tensor defined by its conmutator, which will be related to the ones computed just with the usual metric $g_{\mu\nu}$ in a fancy way. We consign details of those computations to the appendix but just let us recall the final result for the Weyl curvature scalar in terms of the usual one together with the spurion field, which is
\begin{align}
{\cal R}=R-2\sqrt{\frac{n-1}{n-2}}\nabla^{2}\sigma-(\nabla\sigma)^{2}.
\end{align}

The success of this construct is that, via an integration by parts, it corresponds exactly with the Lagrangian density of the scalar-tensor theory, so the full action  can be rewritten in a manifestly Weyl invariant way as
\begin{align}\label{action}
S=\int d^{n}x\sqrt{G}\; {\cal R} = \int d^{n}x\sqrt{g}~e^{-\sqrt{\frac{n-2}{n-1}}\sigma}\left(R+(\nabla\sigma)^{2}\right)
\end{align}
and this shows clearly how the Weyl invariant scalar-tensor theory is just a completion of the usual Einstein-Hilbert theory in order to have pointwise conformal invariance through the gauge field $W_{\mu}$.

It is also interesting to check what the partial gauge fixing $C=1$ means with respect to  conformal invariance. From the equation[\ref{fixing}], we can see that it reduces to just $G=1$, which is exactly the unimodularity condition that we also imposed in the Einstein-Hilbert Lagrangian to define the Unimodular theory. This clearly shows that this theory, at least at the classical level, is no more than a common partially gauge fixed sector of both General Relativity and Conformal Gravity, corresponding to those physical systems that, maintaining conformal invariance (which implies the impossibility of adding a cosmological constant term to the Lagrangian) have general coordinate transformations invariance reduced to area preserving diffeomorphisms only.

This statement has also another useful implication, which is that when written through the conformally invariant metric $G_{\mu\nu}$, the background field expansion of the action is straightforward and identical to the expansion of the Einstein-Hilbert Lagrangian with the added step of changing all geometrical quantities by the ones constructed through $W_{\mu}$. This is easily understood since the covariant structure is the same in both cases and the only difference is the adding of conformal invariance.

\section{One-loop computation}\label{one-loop}
Our goal in this work was to determine  whether the conformal invariance of unimodular gravity was broken by quantum corrections in the form of a trace anomaly.
\par
 There is a general issue of consistency here. 
 \par
 When computing anomalies, the problem is usually reduced to a theory propagating in a background (non-dynamical) gravitational field. This gives rise to the computation of determinants that depend upon the background metric.  What we are doing in this paper is slightly different, in the sense that we are considering the gravitational field as a dynamical entity, and computing its one loop effects. It is a fact that the Einstein-Hilbert Lagrangian is non renormalizable. This has been shown to be the case also for the unimodular variants, as studied in \cite{AlvarezF}. The consistency of our approach is then not guaranteed.   The meaning of our result is then rather that no obvious inconsistency appears when considering the theory to one loop order. This fact alone is highly nontrivial.
 \par
As it is explained in Appendix \ref{B}, the computation of the conformal anomaly can be reduced to the calculation of  the $n=d$ Schwinger-de Witt coefficient in the expansion of the heat kernel corresponding to the quadratic differential operator of the effective action for quantum fluctuations. However, when the expression [\ref{action}] is taken into account, things are easier, since what the heat kernel expansion states is that the conformal (or trace) anomaly is 
\be
-\int d(vol)T = \lambda a_{d}
\ee
where $-\lambda$ is the conformal weight of the corresponding second order operator, $T\equiv T_{\m\n}~g^{\m\n}$ is the trace of the one-loop energy momentum tensor, and $a_d$ is certain coefficient in the expansion of the heat kernel of the operator of quadratic fluctuations as given in the Appendix, formula (B.8).So if we are dealing with pointwise conformal operators in our Lagrangian, this vanishes identically and computing the Schwinger-de Witt coefficient is not necessary. And, recalling what we proved before, this is exactly the situation we are dealing with ,so we should expect the conformal anomaly to cancel in this theory. However, let allow us to be more explicit and compute the counterterm exactly by recalling that the full action of the unimodular theory in the scalar-tensor description was written in a manifestly conformally (Weyl) invariant way, namely

\begin{align}
S=-M^{n-2}\int d^{n}x\sqrt{G}\; {\cal R}=-M^{n-2}\int d^{n}x\sqrt{g}~e^{-\sqrt{\frac{n-2}{n-1}}\sigma}\left(R+(\nabla\sigma)^{2}\right)
\end{align}
Therefore, performing a background field expansion (which has been discussed in some detail in the second reference of \cite{AlvarezF})
\begin{align}
&g_{\mu\nu}\equiv\bar{g}_{\mu\nu}+h_{\m\n}\\
\nonumber&\sigma\equiv\bar{\sigma}+\sigma
\end{align}
provided with the (often dubbed classical) conformal transformations
\begin{align}
\delta_{C}\bar{g}_{\mu\nu}&=2\omega(x)\bar{g}_{\mu\nu}\\
\nonumber \delta_{C}h_{\mu\nu}&=2\omega(x)h_{\mu\nu}\\
\nonumber \delta_{C}\bar{\sigma}&=\sqrt{(n-1)(n-2)}\;\omega\\
\nonumber \delta_{C}\sigma&=0
\end{align}
we can reconstruct again the conformal invariant structure, this time at the linear level, by expanding all quantities in
 the same way as we did in section \ref{invariancia}, but using this time the background field, so we will denote everything computed this way by adding a bar over it. The fact that the variation of the dinamical spurion $\sigma$ vanishes 
 means that all the expressions of Weyl invariant geometrical quantities will be identical to the ones at the non-linear level by just replacing the full field $\sigma$ by the background \,\footnote{It is worth remarking that doing this, the covariant derivative of the gravitational fluctuation $h_{\mu\nu}$, which is a tensor of conformal weight $\lambda=-2$, transforms as another conformal tensor of the same weight.} one $\bar{\sigma}$ 
 and since all these changes can be encoded, as we showed before, into a conformal rescaling of the metric, this implies that the perturbative expansion of this action will match the well-known one of Einstein-Hilbert Lagrangian with just the corresponding change of metric and operators done at every step. So doing it and taking care of fixing the gauge in a conformally (Weyl) background invariant way \footnote{The best option, taking into account that our goal is to compute the gravitational one-loop counterterm, is generalizing the harmonic gauge to $\bar{D}_{\mu}h^{\mu\nu}=0$.}, we are done. 

The background (zeroth order) term reads simply
\begin{align}
\bar{S}=-M^{n-2}\int d^{n}x\; \sqrt{\bar{g}}\;e^{-\sqrt{\frac{n-2}{n-1}}\bar{\sigma}}\bar{{\cal R}}.
\end{align}

On the other hand, the linear terms that have to cancel in order to ensure absence of tadpoles are
\begin{align}
S_{\sigma}&=-M^{n-2}\int d^{n}x\; \sqrt{\bar{g}}e^{-\sqrt{\frac{n-2}{n-1}}\bar{\sigma}}\sqrt{\frac{n-2}{n-1}}\;\bar{{\cal R}}\sigma\\
S_{h}&=M^{n-2}\int d^{n}x\;\sqrt{\bar{g}}e^{-\sqrt{\frac{n-2}{n-1}}\bar{\sigma}}\sqrt{\frac{n-2}{n-1}}\;\bar{\epsilon}_{\mu\nu}h^{\mu\nu}
\end{align}
where $\bar{\epsilon}_{\mu\nu}$ is the background Einstein tensor and we have performed a convenient partial integration in the gravitational fluctuation action. The linear equations of motion for the background metric are then encoded into these linear terms and read 
\begin{align}
-\bar{R}_{\alpha\beta}+\frac{1}{2}\bar{R}\bar{g}_{\alpha\beta}=\bar{\nabla}_{\alpha}\bar{\sigma}\bar{\nabla}_{\beta}\bar{\sigma}-\frac{1}{2}\left(\bar{\nabla}\bar{\sigma}\right)^{2}\bar{g}_{\alpha\beta}.
\end{align}
The  trace of the above implies directly $\bar{R}=-\left(\bar{\nabla}\bar{\sigma}\right)^{2}$ and on the other hand, the geometrical Bianchi identities demand that
\begin{align}
0=\bar{\nabla}^{\alpha}\left( -\bar{R}_{\alpha\beta}+\frac{1}{2}\bar{R}\bar{g}_{\alpha\beta}\right)=\bar{\nabla}^{2}\bar{\sigma}\bar{\nabla}_{\beta}\bar{\sigma}=\frac{1}{2}\sqrt{\frac{n-2}{n-1}}\;\bar{\nabla}_{\beta}\bar{\sigma}\left( \left( \bar{\nabla}\bar{\sigma}\right)^{2}-\bar{R}\right).
\end{align}

Altogether they imply $\bar{R}=\left( \bar{\nabla}\bar{\sigma}\right)^{2}=\bar{\nabla}^{2}\bar{\sigma}=0=\bar{{\cal R}}$, which, as with Einstein equations, is no more than a consequence of the background equations of motion once we take account of the substitution of operators by conformal ones that we were discussing. 

Finally and as we argued, the second order term has to be the same as in the expansion of the Einstein-Hilbert Lagrangian, where at each step the substitution 
\begin{align}
\bar{g}_{\mu\nu}\rightarrow \bar{G}_{\alpha\beta}=e^{-\frac{2}{\sqrt{(n-2)(n-1)}}\bar{\sigma}}\bar{g}_{\alpha\beta}
\end{align}
 is made, which implies also subtituing all derivatives by the background Weyl invariant one $\bar{D}_{\mu}$.

  Thus
\begin{align}
\nonumber S_{h^{2}}=-M^{n-2}\int d^{n}x \sqrt{\bar{G}}\left[\frac{1}{4}\bar{D}^{\mu}H\bar{D}_{\mu}H-\frac{1}{2}\bar{D}_{\mu}H\bar{D}^{\rho}H^{\mu}_{\rho}+\frac{1}{2}\bar{D}_{\mu}H^{\mu\rho}\bar{D}_{\nu}H^{\nu}_{\rho}-\right.\\
\left. -\frac{1}{4}\bar{D}_{\mu}H^{\nu\rho}\bar{D}^{\mu}H_{\nu\rho}-\bar{{\cal R}}_{\nu\beta}H^{\beta}_{\alpha}H^{\nu\alpha}+\frac{1}{2}h\bar{{\cal R}}_{\alpha\beta}H^{\alpha\beta}-\frac{\bar{{\cal R}}}{2}\left( \frac{H^{2}}{4}-\frac{1}{2}H^{\alpha\beta}H_{\alpha\beta}\right)\right]
\end{align}
 
 where $H_{\mu\nu}$ is the graviton fluctuation of the rescaled metric $G_{\mu\nu}$, corresponding to
\begin{align}
H_{\mu\nu}=e^{-\frac{2}{\sqrt{(n-2)(n-1)}}\bar{\sigma}}\left(h_{\mu\nu} -\frac{2\sigma \bar{g}_{\mu\nu}}{\sqrt{(n-2)(n-1)}}\right).
\end{align}

This in turn means that (provided that the corresponding conformal harmonic gauge fixing is used) the counterterm is simply given in terms of the t'Hooft-Veltman \cite{Hooft} counterterm by performing the same  operator substitution we were doing formerly

\begin{align}
\nonumber S_{c}&=\frac{1}{8\pi^{2}(n-4)}\frac{203}{80}\int d^{n}x\; \sqrt{\bar{G}}\;\bar{{\cal R}}^{2}=\\
&=\frac{1}{8\pi^{2}(n-4)}\frac{203}{80}\int d^{n}x\; \sqrt{\bar{g}}\;e^{-\sqrt{\frac{n-2}{n-1}}\bar{\sigma}}\left(\bar{R}-2\sqrt{\frac{n-1}{n-2}}\;\bar{\nabla}^{2}\bar{\sigma}-\left( \bar{\nabla}\bar{\sigma}\right)^{2}\right)^{2}
\end{align}
which is manifestly pointwise conformally invariant and also it vanishes on-shell when background equations of motion are taken into account. This  is in accord with the naive fact that the conformal anomaly should vanish owing to the manifest conformal invariance of the action.

\par

The inclusion of  non-interacting conformal matter does not change the situation. 
For example, a scalar field interacts with the gravitational field according to
\be
S_{matt}\equiv \int d^n x{1\over 2}~g_E^{\m\n}\nabla_\m\Phi\nabla_\n\Phi=\int d^n x~g^{1\over n}{1\over 2}~g^{\m\n}\nabla_\m\Phi\nabla_\n\Phi
\ee

Once embedded in a diffeomorphism invariant theory, the action principle reads
\be
S=\int d^n x \sqrt{\bg}~e^{-\sqrt{n-2\over n-1}\bs}~{1\over 2}\bg^{\m\n} \bn_\m \Phi\bn_\n\Phi
\ee
and given the transformation of $\bs$, it is plain to check that the operator
\be
\Delta f\equiv \bn_\m\left(\sqrt{\bg}~e^{-\sqrt{n-2\over n-1}\bs}~\bg^{\m\n} \bn_\m f\right)
\ee
is conformally invariant.

\par

\section{Conclusions.}
It has been shown that the conformal invariance of unimodular gravity survives quantum corrections, even in the presence of scalar conformal matter. This result is a consequence of the fact that the corresponding operator governing quadratic fluctuations around an arbitrary background is manifestly conformal invariant (vanishing conformal weight).
\par
Another way of looking at this result is through the computation of the counterterm, which is quite simply determined from the standard 't Hooft-Veltman counterterm. This counterterm is Weyl invariant for any dimension, id est, its variation vanishes as opposed to being proportional to $n-4$. It actually vanishes on shell, once the background equations of motion are used. The fact that the conformal anomaly should vanish for unimodular gravity was already conjectured by Blas in his Ph.D. thesis work \cite{Blas}.
\par
The physical situation is not unlike the gauge current in a vectorlike gauge theory, where it is also quite plain that no anomaly is present.
\par
As a general remark, the unimodular theory can be understood as a certain truncation of the full Einstein-Hilbert theory, where in a certain frame (the Einstein frame) the metric tensor is unimodular (with determinant equal to one). Our result is compatible with the idea that the corresponding restriction at the quantum level (i.e. in the functional integral) is  consistent as well.
\newpage
\appendix
\section{Weyl covariant curvature}

Once the Weyl covariant derivative defined through the gauge field $W_{\mu}$ is constructed, geometrical quantities can be computed. To start with, the commutator of two of such  derivatives defines a curvature through Ricci's identity (and is independent of the conformal weight of the tensor acted upon, so the in appearance arbitrary term  $\lambda W_{\mu}T$ does not cause any contradiction and indeed it is needed to ensure  that the derivative of the metric vanishes)
\bea
&&{\cR}_{\m\n\r\s}=R_{\m\n\r\s}-g_{\m\r}\left(\nabla_\n W_\s+W_\n W_\s\right)-g_{\m\s}\left(\nabla_\n W_\s + W_\n W_\r\right)\\
&&+g_{\n\r}\left(\nabla_\m W_\s + W_\m W_\s \right)+g_{\n\s}\left(\nabla_\m W_\r + W_\m W_\r \right)+\left(\nabla_\l W^\l\right)^2\left(g_{\m\s}g_{\n\r}-g_{\m\r}g_{\n\s}\right)=\nonumber\\
&&R_{\m\n\r\s}+g_{\m\r}\left(\nabla_\n \nabla_\s\s+\nabla_\n\s \nabla_\s\s\right)-g_{\m\s}\left(\nabla_\n \nabla_\s \s+ \nabla_\n\s \nabla_\r\s\right)\nonumber\\
&&-g_{\n\r}\left(\nabla_\m \nabla_\s\s + \nabla_\m\s \nabla_\s\s \right)+g_{\n\s}\left(\nabla_\m \nabla_\r\s + \nabla_\m\s \nabla_\r\s \right)+\left(\nabla\s\right)^2\left(g_{\m\s}g_{\n\r}-g_{\m\r}g_{\n\s}\right).\nonumber
\eea

It is easy to realize that, defining a new metric by a conformal rescaling $G_{\alpha\beta}=e^{-\frac{2\sigma}{\sqrt{(n-2)(n-1)}}}g_{\alpha\beta}$, what we have is
\bea
{\cR}_{\m\n\r\s}&&=e^{{2\over \sqrt{(n-1)(n-2)}}\s}~R_{\m\n\r\s}\left[g_{\a\b}~ e^{-{2\over \sqrt{(n-1)(n-2)}}\s}\right]=\left( \frac{G}{g}\right)^{1/n}~R_{\m\n\r\s}\left[G_{\a\b}\right]
\eea
which corresponds to the usual Riemman tensor that we would compute using the metric $G_{\a\b}$ with a prefactor $(G/g)^{1/n}$ whose origin is to ensure pointwise conformal invariance. Accordingly
\bea
{\cR}_{\m\n}&&=R_{\m\n}+(n-2)\left(\nabla_\m W_\n+ W_\m W_\n\right)+g_{\m\n}\left(\nabla_\l W^\l -(n-2) W_\l W^\l\right)=\nonumber\\
&&=R_{\m\n}-\sqrt{n-2\over n-1}\nabla_\m \nabla_\n\s+ {1\over n-1}\nabla_\m\s \nabla_\n\s-\nonumber\\
&&-g_{\m\n}\left({1\over\sqrt{(n-1)(n-2)}}\nabla^2\s +{1\over n-1}\nabla_\l\s \nabla^\l\s\right).\nonumber
\eea

And this Ricci tensor has also got a quite simple interpretation
\be
{\cR}_{\m\n}=e^{{2\over \sqrt{(n-1)(n-2)}}\s}~R_{\m\n}\left[g_{\a\b}~ e^{-{2\over \sqrt{(n-1)(n-2)}}\s}\right]=\left( \frac{G}{g}\right)^{1/n}~R_{\m\n}\left[G_{\a\b}\right]
\ee
manifestly conformal invariant under
\bea
&&g_{\m\n}\rightarrow \Omega^2 g_{\m\n}\nonumber\\
&&e^{-{2\over \sqrt{(n-1)(n-2)}}\s}\rightarrow \Omega^{-2}e^{-{2\over \sqrt{(n-1)(n-2)}}\s}.
\eea

From this, the curvature scalar is straightforward and inherits the same interpretation
\bea
&&{\cR}=R + 2 \left(n-1\right) \nabla_\l W^\l -\left(n-2\right)\left(n-1\right) W_\l W^\l=\nonumber\\
&&=R- 2 \sqrt{n-1\over n-2} \nabla^2 \s - \left(\nabla\s\right)^2=\left( \frac{G}{g}\right)^{1/n}~R\left[G_{\a\b}\right].
\eea

From all this, the Einstein tensor results to be
\be
{\cal E_{\m\n}} = R_{\m\n}-{1\over 2} R g_{\m\n}+{1\over n-1}\nabla_\m\s\nabla_\n\s+\sqrt{(n-1)(n-2)}\nabla^2\s g_{\m\n}+{n-3\over 2(n-1)}(\nabla\s)^2 g_{\m\n}.
\ee
\par

Finally, taking into account that the measure 
\be
\sqrt{g} ~e^{-{n\over\sqrt{(n-1)(n-2)}}\s}~d^n x = \sqrt{G}\left( \frac{G}{g}\right)^{-1/n}
\ee
is conformal invariant, the only dimension two pointwise invariant operator is
\be
\int d^n x~\sqrt{g} e^{-{n\over\sqrt{(n-1)(n-2)}}\s}{R}=\int d^n x \sqrt{g}~e^{-\sqrt{n-2\over n-1}\s}\left(R+\left(\nabla\s\right)^2\right)
\ee
 and after integration by parts, the full action can then be  written as
\be
S= \int d^n x~\sqrt{g} e^{-{n\over\sqrt{(n-1)(n-2)}}\s}{\cal R}=\int d^n x~\sqrt{G}\;\cal R
\ee
where the factors $G/g$ cancel exactly and show how dynamics can be obtained from the metric $G_{\a\b}$ even if it does not encode all information about the nature of the Weyl covariant derivative (explicitely, it  knows nothing about the $\lambda W_{\m}T$ term of the derivative).
\par

At the linear level, the conformal classical (or background) transformations are
\bea
&&\delta_{C}\bar{g}_{\mu\nu}=2\omega(x)\bar{g}_{\mu\nu}\nonumber\\
 &&\delta_{C}h_{\mu\nu}=2\omega(x)h_{\mu\nu}\nonumber\\		
 &&\delta_{C}\bar{\sigma}=\sqrt{(n-1)(n-2)}\;\omega\nonumber\\
  &&\delta_{C}\sigma=0
\eea
and since they vanish for the spurion field fluctuation, this means that all the geometrical construct we just did in this appendix can be redone on the background field expansion as well just by replacing $\sigma$ by $\bar{\sigma}$.

\section{Conformal anomaly}\label{B}
It is well known that one loop computations are equivalent to the calculation of functional determinants. One of the simplest definitions of the determinant of an operator is through the $\zeta$-function technique \cite{Hawking}. We shall follow conventions as in \cite{AlvarezFV}. Given a differential operator of the general form
\be
\Delta\equiv-D_\m D^\m +Y
\ee
with $D_\m\equiv \pd_\m +X_\m$, we assume that the elliptic operator $\Delta$ enjoys eigenvalues $\l_n$
\be
\Delta \phi_n=\l_n \phi_n
\ee
normalized in such a way that
\be
\int d^n x~\sqrt{g}~\phi_i~\phi_j=\d_{ij}.
\ee

Now the {\em heat kernel} is formally defined as
\be
K(\t)\equiv e^{-\t \Delta}
\ee
and its action on functions reads
\be
(Kf)(x)=\int~ d(vol)_y~ K(x,y;\t) ~f(y).
\ee
The ultraviolet (UV) behavior is controlled by the short time Schwinger-de Witt expansion which reads
\be
K(x,y;\t)=K_0(x,y;\t)~\sum_{p=0} b_{2p}\t^p
\ee
where for instance the flat space kernel reads
\be
K_0(x,y;\t)={1\over (4\pi\t)^{n\over 2}}~e^{-{(x-y)^2\over 4\t}}.
\ee
The integrated quantity $Y(\t,f)\equiv\text{tr}~(K f)$ also enjoys a corresponding short time expansion
\be
Y(\t,f)=\sum_{k=0}\t^{k-n\over 2}~ a_k(f).
\ee
The trace in the preceding formulas involves spacetime integration as well as sum over all finite rank indices.  Sometimes one simply writes $Y(\t)\equiv Y(\t,1)$. 
\par

The {\em zeta function} is defined as
\be
\Gamma(s)\zeta(s)=\int_0^\infty dt~t^{s-1}~Y(t)=\sum_n \l_n^{-s}
\ee
where the second equality is even more formal than the first one. 

The determinant of the differential operator is then defined \cite{Hawking}  as
\be
\text{det}~\Delta\equiv \prod_n \l_n\equiv e^{-\zeta^\prime(0)}
\ee

Now assume that we have a quantum field theory that we dimensionally regularize, id est, we make $n=d+\e$, where $d$ is the physical dimension (for example $d=4$), then, at the one-loop level, there is a divergent piece in the effective action
\be
W_\infty=-{1\over 2}\left.\text{log~det}\Delta\right|_\infty=-\m^\e~{a_d\over \e}.
\ee
 On the other hand, when performing a {\em rigid} Weyl transformation on the spacetime metric
\be
\widetilde{g}_{\m\n}=\Omega^2~g_{\m\n}=\left(1+2 \omega\right)g_{\m\n}
\ee
the eigenvalues of the operator transform in a definite manner which coincide with the {\em conformal weight} $\l$ of the operator.
\be
\widetilde{\l_n}\equiv \Omega^{-\l}~\l_n.
\ee
Usually the conformal weight  is just the mass dimension of the operator in the sense of dimensional analysis.
\par
According to Branson \cite{Erdmenger} a {\em conformal covariant} operator $D$ transforms under {\em local} (not only rigid)  Weyl transformations in such a way that there exist two numbers $(a,b)$ such that the Weyl rescaled operator is given by
\be
\widetilde{D}\phi=\Omega^{-b}~D\left(\Omega^a \phi\right).
\ee
It follows that  that the new eigenfunctions are given by 
\be
\widetilde{\phi}_n=\Omega^{-a}\phi_n
\ee
 and the new eigenvalues by
\be
\widetilde{\l}_n=\Omega^{-b} \l_n.
\ee 
The archetype of such operators is the conformal laplacian
\be
\Delta_c\equiv \Delta-{1\over 4}{n-2\over n-1} R
\ee
which is such that
\be
\overline{\Delta}_c \left(\Omega^{-{n-2\over 2}}~\phi\right)=\Omega^{-{n+2\over 2}}\Delta\phi.
\ee
There are no known  diffeomorphisms invariant operators built out of the metric alone with $b=0$.

\par
In the case of the standard scalar laplacian,
\be
\Delta\equiv \nabla^2\equiv{1\over \sqrt{g}}\pd_\m\left(g^{\m\n}\sqrt{g}\pd_\n\right)
\ee
 the conformal weight coindices with its mass dimension, $\l=2$.
\par
The new zeta function after the Weyl transformation is given in general by
\be
\widetilde{\zeta}(s)=\Omega^{D s}~\zeta(s)
\ee
so that the determinant defined through the $\zeta$-function scales as
\be
\text{det}~\widetilde{\Delta}=\Omega^{-\l\zeta(0)}~\text{det}~\Delta
\ee
and this modifies correspondingly the effective action
\be
\widetilde{W}=W+~\l~\omega~\zeta(0).
\ee
The energy-momentum tensor is {\em defined} in such a way that under a general variation of the metric the variation of the effective action reads  
\be
\d W\equiv {1\over 2}\int d(vol)_x T_{\m\n}\d g^{\m\n}
\ee
which in the particular case that this variation is proportional to the metric tensor itself (like in a conformal transformation at the lineal level), $\d g^{\m\n}=-2 \omega g^{\m\n}$ yields the integrated trece of the energy-momentum tensor
\be
\d W=-\int d(vol) \omega T.
\ee
Conformal invariance in the above sense then means that the energy-momentum tensor must be traceless. When quantum corrections are taken into account, it follows  that
\be
-\int d(vol)  T= \l~\zeta(0).
\ee

It is not difficult to show that
\be
\zeta(0)\equiv \lim_{s\rightarrow 0} s \int_0^\infty~ dt~ t^{s-1} Y(t)=\lim_{s\rightarrow 0} s \int_0^1~ dt~ t^{s-1} Y(t)=a_d
\ee
where $n=d$ is the specific value of the spacetime dimension. The conformal anomaly is usually then written as
\be
-\int d(vol)  T = \l a_{d}.
\ee
 The Schwinger-de Witt coefficient corresponding to the physical dimension, $n=d$  precisely coincides with the divergent part of the effective action when computed in dimensional regularization as indicated above. This means that in order to compute the one loop conformal anomaly in many cases it is enough to compute the corresponding counterterm.
\par
This argument shows clearly that when the conformal weight of the operator of interest vanishes, $\l=0$ all eigenvalues remain invariant and there is no conformal anomaly for determinants defined through the zeta function. In our case this will follow from the manifest Weyl invariance of the construction of the operator at all steps. This conformal invariance in turn in inherited from the mother theory which enjoys invariance under area preserving diffeomorphisms only. This is the origin of the background dilaton $\bs$ of gravitational origin, essential in our approach.

\newpage
\section*{Acknowledgments}
We have enjoyed many discussions with Luis Alvarez-Gaum\'e. This work has been partially supported by the European Union FP7  ITN INVISIBLES (Marie Curie Actions, PITN- GA-2011- 289442)and (HPRN-CT-200-00148) as well as by FPA2009-09017 (DGI del MCyT, Spain) and S2009ESP-1473 (CA Madrid). M.H. acknowledges a "Campus de Excelencia"  grant from the Departamento de F\'{\i}sica Te\'orica of the UAM. The authors acknowledge the support of the Spanish MINECOÕs ÒCentro de Excelencia Severo OchoaÓ Programme under grant  SEV-2012-0249.

\appendix

\end{document}